

\magnification=\magstep1
\overfullrule=0pt

\def\intdtq{\int{{d^3q}\over{(2\pi)^3}}}

\def\frac#1#2{{#1\over#2}}

\line{ }
\hfill HD-TVP-94-13
\vglue 1.0in
\centerline{{\bf
Hadronization cross--sections}}
\centerline{{\bf at the chiral phase transition of a quark plasma}
\footnote{$^1$}{Supported in part by the Federal Minister for
Research and Technology (BMFT) grant number 06 HD 729, and the Deutsche
Forschungsgemeinschaft, grant number Hu 233/4-2.}}
\medskip
\centerline{ J. H\"ufner, S.P. Klevansky, E.Quack
and P. Zhuang\footnote{$^2$}{A.v. Humboldt fellow. On leave of absence from
             Huazhong Normal University, Wuhan, China. }}
\smallskip
\centerline{Institut f\"ur Theoretische Physik, Universit\"at
Heidelberg,}
\centerline{Philosophenweg 19, 69120 Heidelberg, Germany.}
\smallskip
\centerline{June 1994}
\smallskip
\vglue 1.0in
\noindent ABSTRACT:   Hadronization at finite temperature $T$ is discussed
in the framework of the Nambu--Jona-Lasinio model. The differential
cross-section for the conversion of a quark--antiquark pair into two pions
to first order in a $1/N_c$ expansion is calculated as a function of the
c.m.~energy $s$ and temperature $T$.
In particular, approaching the temperature $T_c$ of the chiral phase
transition, the hadronization cross-section diverges like
$\ln |1-{T\over T_c}|$.

\vskip 0.2 in
\noindent
Submitted to Phys. Lett. B

\vglue 0.1in
\vfill\eject
\baselineskip=20pt
\noindent

Our current understanding of the hadronic world via quantum chromodynamics
(QCD) appears to be characterized by (a) confinement, in that no free quarks
and gluons are observed, and (b) broken chiral symmetry, manifesting itself in
a non-vanishing value of the
quark condensate
$<\bar \psi \psi >$ and in a small pion mass.
In particular, chiral symmetry is essential in determining the static
properties of the light hadrons. In ultra-relativistic heavy ion collisions,
nuclear matter is heated up to a degree where one might penetrate
into the region of both a chirally symmetric
and a deconfined phase of matter, at a transition temperature in the region
$T_c \sim 100-200$ MeV [1,2]. In the cooling process
the deconfined partons
will recombine into the observed hadrons, which are
mainly pions. It is thus important to understand how
hadronization rates for processes such as
the conversion of a quark--antiquark pair into two pions,
$q \bar q \rightarrow 2\pi$, behave
as a function of the c.m.~energy $\sqrt{s}$ and the temperature $T$,
and especially what the influence is of a phase
transition on these rates.

In this paper, we study the hadronization rates for $q \bar q \rightarrow 2\pi$
in a model system that displays chiral symmetry, the flavor $SU(2)$
Nambu--Jona-Lasinio (NJL) model [3,4]. We are aware of the deficiencies of the
model---
the lack of confinement and non-renormalizability of the effective 4-fermion
interaction, but also of its strengths---the transparent description of the
chiral
phase transition at critical temperature $T_c$, with the attendant binding of
quarks into mesons. For
temperatures $T\ge T_c$, no hadrons exist in the model, and one has a system
of interacting quarks with mass $m=0$ in the chiral limit. For $T\le T_c$,
in the domain of broken symmetry, a quark-meson plasma is formed with $m\ne 0$,
pion mass $m_{\pi}=0$ and
sigma mass $m_{\sigma}=2m$. While the situation for $T>T_c$ is fairly
realistic (except for the absence of gluons), the coexistence of quarks and
mesons below $T_c$ is an artifact of the NJL-model. Yet with due caution,
one may
still extract meaningful, physical information from the NJL model about
hadronization in the vicinity of the phase transition itself.

The NJL Lagrange density determining the chiral dynamics of quarks and mesons
in flavor $SU(2)$ is
$$
{\cal L}(x) = \bar
\psi(x)(i\not\partial - m_0)\psi(x) + G[(\bar\psi(x) \psi(x))^2 +
(\bar
\psi(x) i \gamma_5\vec\tau \psi(x))^2] ,
\eqno(1)
$$
where G is a coupling constant of dimension [Mass]$^{-2}$, and $m_0$ is the
current quark mass. In this paper, we work for the most part in the chiral
limit
$m_0=0$, where the system displays a phase transition at a critical
temperature $T_c$. For simplicity we restrict ourselves to $\mu=0$, i.e.~zero
baryon density, which corresponds to the central rapidity region of heavy ion
collisions. A three dimensional regulator $\Lambda$ will be
imposed on all divergent quantities. Since the NJL model is a strong--coupling
theory, perturbation theory is inapplicable, and we require a selection
procedure for the relevant Feynman graphs. We
choose the $1/N_c$ expansion [5-7], and work to lowest order. To $O(1/N_c)$,
the relevant T-matrix amplitudes for the reaction $q\bar q\rightarrow 2\pi$
are sketched in Fig.1. Figure 1(a) depicts
the s-channel amplitude $T^{(s)}$ in which an intermediate $\sigma$-meson
is produced,
while Figure 1(b) displays quark exchange in the t-channel, the
amplitude being denoted by $T^{(t)}$.
It is understood
that the crossed diagrams are included for a process
in which the final state particles are identical, e.g. for $u\bar u \rightarrow
2\pi_0$.

Using an obvious notation for the T-matrix amplitudes, one has
$$
T^{(s)}(k,p_3;T) = \bar v_2u_1{2G\over 1-2G\Pi_\sigma (k;T)}g_{\pi}^2
A_{\sigma\pi\pi}(k,p_3;T)\delta_{c_1,c_2} ,
\eqno (2)
$$
where the symbol $\delta_{c_1,c_2}$ refers to the color degree of freedom and
$A_{\sigma\pi\pi}$ represents the amplitude of the triangle vertex
$\sigma \rightarrow \pi \pi$. In
terms of the intermediate $\sigma$-meson four--momentum $k$ and the
four--momentum $p_3$ for one external pion, $A_{\sigma\pi\pi}$ is obtained by
analytic continuation of
$$\eqalign{
& A_{\sigma\pi\pi}(i\nu_l,\vec k;i\mu_m,\vec p_3;T)= \cr
& Tr{1\over \beta}\sum_n e^{i\omega_n \eta}\intdtq
S(i\omega_n,\vec q)i\gamma_5\vec \tau S(i\omega_n+i\mu_m,\vec q+\vec p_3)
i\gamma_5\vec \tau S(i\omega_n+i\nu_l,\vec q+\vec k)  .\cr}
\eqno (3)
$$
Here $\omega_n=(2n+1)\pi/\beta, n=0,\pm1,\pm2,
...$ and $\mu_m=2m\pi/\beta, \nu_l=2l\pi/\beta, m,l=0,\pm1,\pm2,...$ are
fermionic and bosonic Matsubara frequencies respectively. The quark propagator
is denoted by $S(i\omega_n,\vec q)
=[\gamma_0(i\omega_n)-\vec \gamma \vec q +m]/[(i\omega_n)^2-E_q^2]$, with
$E_q^2=
\vec q^2+m^2$, where $m=m(T)$ is the
dynamically generated quark mass, to be calculated using the usual gap
equation in the Hartree approximation [4]. In Eq.(3), $Tr$ refers
to the trace over color, flavor and spinor indices. Finally, $\Pi_\sigma(k;T)$
in Eq.(2) is the standard mesonic polarization function for the
$\sigma$--meson [4], and $g_\pi$ is the
pion-quark coupling strength, determined in the model via [4]
$$
g_\pi^{-2}(T) = {\partial \Pi_\pi(k_0,\vec 0;T)\over \partial k_0^2}|_{
k_0^2=m_\pi^2} .
\eqno (4)
$$
The indices 1 and 2 of the quark and antiquark spinors $u$ and $\bar v$
in Eq.(2) refer
to both momenta and helicity.

After evaluation of the Matsubara sum in Eq.(3), it is useful, in the analytic
continuation to real variables, to move to the center of mass frame of the
incident $q\bar q$ pair. While the differential cross-section of any two--body
reaction $d\sigma(s,t)/dt$ is invariant under a Lorentz-boost, this is no
longer so in a heat bath. Here the cross section not only depends on the
temperature $T$, but also on the c.m.~velocity of the initial pair
with respect to the heat bath. By assuming
the $q\bar q$ system to be at rest in the heat bath, we avoid this
complication, but also restrict ourselves to only part of the solution.
Then one has $i\nu_l\rightarrow k_0=\sqrt{s}, \;
i\mu_m\rightarrow p_3^0=\sqrt {s}/2, \; |\vec p_3|=\sqrt{s/4-m_\pi^2}$ and
$p_3k=s/2$, leading to the form
$$
A_{\sigma\pi\pi}(s;T)=2mN_cN_f\intdtq {f(E_q)-f(-E_q)\over E_q}
{2sE_q^2-s^2/2-8(\vec q\vec p_3)^2+(4m_\pi^2+2s)\vec q\vec p_3 \over
(s-4E_q^2)((m_\pi^2-2\vec q\vec p_3)^2-sE_q^2)}
\eqno(5)
$$
where the Fermi function $f(x)=(1+e^{\beta x})^{-1}$.
One sees that the amplitude $A_{\sigma\pi\pi}$ and the T-matrix element
$T^{(s)}$ are simply functions of
the center of mass energy $s=(p_1+p_2)^2$.
Note that the amplitude $A$ in Eq.(5) is proportional to the dynamically
generated quark
mass $m$, a fact that is general to any 3-meson vertex process.
Therefore the s-channel diagrams become unimportant
at the chiral phase transition in the chiral limit. Since $m=0$ for
$T\ge T_c$, one has $A_{\sigma\pi\pi}=0$, while for $T<T_c$, $m_\pi=0$.

For the real and imaginary parts of $A_{\sigma\pi\pi}$, semi-analytic
expressions can be found. Explicitly, one has
$$
Re A_{\sigma\pi\pi}(s;T)=-{mN_cN_f\over 2\pi^2}\int dq (4q^2+{sq\over 2E_q}
\ln|{E_q-q\over E_q+q}|){f(E_q)-f(-E_q)\over E_q(s-4E_q^2)}
\eqno(6)
$$
and
$$
Im A_{\sigma\pi\pi}(s,T)={mN_cN_f\over 8\pi\sqrt{s}}(2\sqrt{s-4m^2}
+\sqrt{s}\ln|{\sqrt{s-4m^2}-\sqrt{s}\over \sqrt{s-4m^2}+\sqrt{s}}|)
(f({\sqrt{s}\over 2})-f(-{\sqrt{s}\over 2})) .
\eqno(7)
$$

Next we turn to the transition amplitude $T^{(t)}$ in the t-channel shown in
Fig.1(b), which reads
$$
T^{(t)}(t;T)=g^2_\pi\bar v_2\gamma_5{\not p_1-\not p_3 +m\over t-m^2}
\gamma _5u_1  ,
\eqno(8)
$$
in which the $m$ and $t=(p_1-p_3)^2$ dependence is made explicit.
This amplitude $T^{(t)}$
forms the dominant contribution for $T\rightarrow T_c$, since, unlike in the
s-channel, it is not directly
proportional to the dynamically generated quark mass. When approaching the
phase transition and $m\rightarrow 0$, one sees that the range of the "force"
due to the exchange of a constituent quark tends to infinity, and we thus
expect that the cross-section becomes infinite at this point.

As an example, we consider in detail the hadronization process
$u\bar u \rightarrow
2\pi_0$. The s-channel crossed graph equals the direct term, while the
t-channel crossed graph gives rise to a distinct exchange term. The
differential cross-section can be written as
$$
{d\sigma_{u\bar u\rightarrow 2\pi_0}\over dt}(s,t;T)=
{1\over 16\pi s(s-4m^2)}\sum_{c,s}{}'|2T^{(s)}(s;T)+T^{(t,dir)}(t;T)
+T^{(t,exc)}(t;T)|^2 .
\eqno (9)
$$
Here the prime on the summation indicates the average over spin
and color degrees of freedom of the quarks in the initial states.
The prefactor multiplying the differential cross-section is
due to the initial flux. The direct contribution to the T--matrix,
$T^{(t,dir)}$, is as given in (9) while the exchange contribution
$T^{(t,exc)}$ can be derived from the u-channel to be
$$
T^{(t,exc)}(t;T)=g^2_\pi\bar v_2\gamma_5{\not p_1-\not p_4 +m\over
u-m^2}\gamma_5 u_1 .
\eqno (10)
$$
The individual contributions to the cross-section are listed in Table 1.

The integrated cross-section multiplied by a factor which takes into account
the Bose-Einstein statistics in form of the distribution function
$f_B(x)=(e^{\beta x}-1)^{-1}$
for the final state pions is
$$
\sigma_{u\bar u\rightarrow 2\pi_0}(s;T)=\int^{t_{max}}_{t_{min}} dt
{d\sigma_{u\bar u\rightarrow 2\pi_0}\over dt}(s,t;T)(1+f_B({\sqrt{s}\over 2}))
^2 \; .
\eqno (11)
$$
The limits $t_{max}$ and $t_{min}$ of the integral
in Eq.(11) are given as
$$
t_{max}=-{s\over 2}+m^2+{1\over 2}\sqrt{s(s-4m^2)}
\eqno (12)
$$
and
$$
t_{min}=-{s\over 2}+m^2 .
\eqno (13)
$$

Before we discuss our numerical results in Figs.2 and 3, we introduce the
hadronization
cross-section into charged pions. The hadronization cross-sections
$u\bar u\rightarrow \pi^+\pi^-$ and $u \bar u\rightarrow \pi^+\pi^0$ are
obtained by simple modifications of the $\pi^0\pi^0$ results. Since
the final state does not involve identical particles, $t_{min}$ is replaced
by
$$
t_{min}=-{s\over 2}+m^2-{1\over 2}\sqrt{s(s-4m^2)}
\eqno (14)
$$
and no crossed diagrams occur. The isospin trace factor leads however to
additional factors of $2$ or $\sqrt{2}$ in the t-channel of the T-matrix
element. One obtains the scattering cross-section for the charged final
states on replacing the primed sum in Eq.(9) by
$$
\sum_{c,s}{}'|T^{(s)}+2T^{(t)}|^2
\eqno (15)
$$
for $u\bar u\rightarrow \pi^+\pi^-$ and
$$
\sum_{c,s}{}'|\sqrt{2}T^{(t)}|^2
\eqno (16)
$$
for $u\bar u\rightarrow \pi^+\pi^0$. We define the total hadronization
cross-section of a single quark as
$$
\sigma_u^{had}=\sigma_{u\bar u\rightarrow \pi^0\pi^0}
              +\sigma_{u\bar u\rightarrow \pi^+\pi^-}
              +\sigma_{u\bar d\rightarrow \pi^+\pi^0} .
\eqno (17)
$$
This is a measure of how quickly a $u$-quark hadronizes from a charge
symmetric plasma. The corresponding cross-section for elastic
scattering [9] is defined as
$$
\sigma_u^{ela}=\sigma_{u \bar u\rightarrow u \bar u}
              +\sigma_{u \bar u\rightarrow d \bar d}
              +\sigma_{u \bar d\rightarrow u \bar d}
              +\sigma_{u u\rightarrow u u}
              +\sigma_{u d\rightarrow u d} .
\eqno (18)
$$

Figures 2 and 3 show our numerical results. In the calculation,
the parameters of the NJL model
have been chosen in the standard way [4], to fit $f_\pi=93$ MeV and
$<\bar \psi \psi>=-(.25 {\rm GeV})^3$ at $T=0$, leading to a value [8]
$G=5.02 {\rm GeV}^{-2}$ and $\Lambda
=.65$ GeV. The critical temperature at which a second order phase
transition occurs is $T_c=.195$ GeV. Because of the use of a cutoff, the
energy range is restricted to $4m^2\le s \le 4(m^2+\Lambda^2)$.

In Fig.2, we show the integrated hadronization cross-section
$\sigma_q^{had} = \sigma_u^{had} = \sigma_d^{had}$
as a function of the c.m.~energy
$s$ for various values of the temperature. The hadronization cross-section
is rather flat, except for a singularity that occurs
at the threshold value $s_{th}
=4m^2$. Note that the threshold is a function of temperature since $m = m(T)$.
While the shape of the s-dependence of $\sigma_q^{had}$ remains similar with
increasing temperature, its magnitude increases dramatically when
approaching the phase transition temperature (about a factor 4 between
T=.15 GeV and .19 GeV). At T=.19 GeV, i.e.~5 MeV below $T_c$, the hadronization
cross-section and the elastic cross-section are of comparable magnitude
(about 5 mb) with a singularity for both at the threshold. There is however
a difference between $\sigma_q^{ela}$ and $\sigma_q^{had}$. While
$\sigma_q^{ela}(s;T)$ remains relatively constant in shape and amplitude, when
going through the phase
transition, $\sigma_q^{had}(s;T)$ displays a singularity at $T=T_c$ for all
values of $s$. This is shown in Fig.3, where $\sigma_q^{had}(s;T)$ is
displayed
as a function of temperature for several values of $s$. Note that there is
no hadronization cross-section for $T>T_c$, since for $T>T_c$, we have
strict deconfinement even in the NJL model.

The singularity structure of $\sigma_q^{had}(s;T)$ which is evident from the
numerical calculations can be summarized by the two statements (considering
always
$T\le T_c$):

\noindent
(i) $\sigma_q^{had}(s;T)\rightarrow \infty $ for $s\rightarrow 4m^2$ and all
$T$,

\noindent
(ii) $\sigma_q^{had}(s;T)\rightarrow \infty $ for $T\rightarrow T_c$ and all
$s$.

Case (i) can be understood to be of kinematical and dynamical origin.
One finds for $s\rightarrow 4m^2$,
$$\eqalign{
& t_{max}\sim -m^2, \cr
& t_{max}-t_{min}\sim \sqrt{s(s-4m^2)}\rightarrow 0 , \cr}
\eqno (19)
$$
so that (for $\sqrt{s-4m^2}<<\sqrt {s}$, i.e. for $T<T_c$)
$$\eqalign{
\sigma^{had}(s)
& \simeq \sqrt{s(s-4m^2)}{d\sigma\over dt}|_{t=t_{max}} \cr
& ={1\over \sqrt{s(s-4m^2)}}(C^{(t)}+C^{(s)}{m^2\over s-4m^2}) ,  \cr}
\eqno (20)
$$
where $C^{(t)}$ and $C^{(s)}$ are smooth functions of $m$ and $s$.
The singularity proportional to $(s-4m^2)^{-1/2}$ is expected,
since the process $q\bar q\rightarrow 2\pi$ is exothermic:
the entrance channel has a threshold energy $\sqrt{s}=2m>0$,
while the exit channel has $\sqrt{s}=0$. This singularity thus appears
for any dynamics. The contribution from $\sigma$-exchange in the
s-channel, the term proportional to $C^{(s)}$, increases the
power of the singularity to $1/(s-4m^2)^{3/2}$. It arises from the s--channel
resonance due to the excitation of the $\sigma$--meson.
However, because of the factor of $m^2$, the effect of the $\sigma$-meson
exchange diminishes when approaching $T_c$.

Case (ii) mentioned above, the singularity for $T\rightarrow T_c$,
has its origin in the chiral dynamics of the model and arises from the
diagram in Fig.~1 b. The leading order behavior from the t-channel exchange
is given by
$$
{d\sigma\over dt}\sim {1\over (t-m^2)} ,
\eqno (21)
$$
which, in combination with the limiting value $t_{max}\simeq -m^4/(16s)
\rightarrow 0$ and $t_{min}=-s$, and the relation [4]
$m=-2G<\bar \psi \psi>\sim (T_c-T)^{1/2}$, leads to the
behavior
$$\eqalign{
\sigma^{had}(s;T)
&\sim \ln|{t_{max}-m^2\over t_{min}-m^2}| \cr
&\sim \ln|<\bar \psi \psi>(T)| \sim \ln|1-{T\over T_c}| \cr}
\eqno (22)
$$
as $T\rightarrow T_c$. Since we work in the chiral limit where the phase
transition is of second order, the hadronization cross-section diverges at
$T=T_c$. Note that this logarithmic singularity is present for all
values of $s$.

In this paper we have investigated the influence of the chiral phase
transition on the hadronization phenomenon in a quark plasma. We have used
a particular model, the NJL model, which has a deconfined phase above
the chiral phase transition temperature $T_c$ but a meson-quark plasma
for $T<T_c$. Although we have calculated the hadronization cross-section
$\sigma_q^{had}(s;T)$ for all temperatures, the results in the neighborhood
of $T_c$ are the most interesting and may be most useful. We find that the
hadronization cross-section is always
strongest for small values of the c.m.~energy of the $q\bar q$ pair, and it
diverges for all values of $s$ when $T\rightarrow T_c$.

We have also considered
the case of a non-vanishing current quark mass $m_0$ ($m_0$ = 5 MeV).
Then the phase
transition is washed out, and so are the singularities in the cross-sections
at the transition temperature. The behavior at the threshold in $s$
(shown in Fig.2) remains, while the logarithmic singularity seen in Fig.3 at
$T=T_c$ vanishes completely.

Although our results are derived in a particular model, we expect them to be
of more general validity. The main conclusions are: (i) The hadronization
cross section in the neighborhood of the phase transition is of the order
of several mb, and strongly depends on the relative momentum of the
quark--antiquark pair. (ii) The dynamics of the
phase transition (here via temperature dependent masses) plays an essential
role and leads to several singularities in the cross-sections.

\vskip 0.2in
\noindent
Acknowledgments

We thank M.Volkov for useful discussions in the course of this work.
One of the authors (P.Z.) wishes to thank the Alexander von Humboldt
Foundation for financial support.

\vskip 0.2in
\noindent
References

\vskip 0.1in
\noindent
[1] See, for example, L.P.Csernai, Introduction to Relativistic Heavy Ion
Collisions, Wiley, New York (1994).

\noindent
[2] Proceedings of the Quark Matter '93, Nucl. Phys. {\bf A 566} (1994).

\noindent
[3] Y. Nambu and G. Jona-Lasinio, Phys. Rev. {\bf 122}
(1961) 345; {\it
ibid.} {\bf 124} (1961) 246.

\noindent
[4] For reviews and general references, see U. Vogl and W.
Weise, Prog.
Part. and Nucl. Phys. {\bf 27} (1991) 195,  S.P. Klevansky, Rev.
Mod.
Phys. {\bf 64} (1992) 649, M.K.Volkov, Phys. Part. Nucl. 24(1993)35,
T.Hatsuda and T. Kunihiro, Phys. Rep. to appear.

\noindent
[5] E. Witten, Nucl. Phys. {\bf B160} (1979) 57.

\noindent
[6] A review on $1/N_c$ expansions is given by S. Coleman, {\it Aspects of
Symmetry, Selected Erice Lectures}, (Cambridge U.P., 1988) p351.

\noindent
[7] In the NJL model, this was first suggested by A.H. Blin, B. Hiller and
M. Schaden, Z. Phys. {\bf A331} (1988) 75. For a detailed treatment see
E.Quack and S.P.Klevansky, Phys. Rev. {\bf C 49} (1994) 3283.

\noindent
[8] J. H\"ufner, S.P. Klevansky, P. Zhuang and H. Voss,
 Ann. Phys.(N.Y.), in press; P. Zhuang, J. H\"ufner and S.P. Klevansky,
 Nucl. Phys. A, in press.

\noindent
[9] P.Zhuang, J.H\"ufner, S.P.Klevansky and L.Neise,
Transport properties of a quark plasma and critical scattering at the
chiral phase transition, Heidelberg preprint HD-TVP-94-09.

\vfill\eject
\vskip 0.2in
\noindent
Table 1

\noindent
Products of the T-matrix amplitudes (in the chiral limit),
which contribute to the hadronization
cross sections, averaged over initial and summed over final spin and color
degrees of freedom.

\noindent
----------------------------------------------------------------------------
---------------------------------
$$
\eqalign{
&\sum_{c,s}{}'|T^{(s)}|^2={2G^2g_\pi^4(s-4m^2)\over N_c}{|A_{\sigma\pi\pi}
(s;T)|^2\over |1-2G\Pi_\sigma(s;T)|^2} \cr
&\sum_{c,s}{}'|T^{(t,dir)}|^2={-g_\pi^4\over 2N_c}[1+{s\over t-m^2}] \cr
&\sum_{c,s}{}'|T^{(t,exc)}|^2={g_\pi^4\over 2N_c}{t-m^2\over m^2-s-t} \cr
&\sum_{c,s}{}'T^{(s)}T^{*(t,dir)}={mGg_\pi^4\over N_c(t-m^2)}(s+2t-2m^2)
Re {A_{\sigma\pi\pi}(s;T)\over 1-2G\Pi_\sigma(s;T)} \cr
&\sum_{c,s}{}'T^{(s)}T^{*(t,exc)}={-mGg_\pi^4\over N_c(m^2-s-t)}(s+2t-2m^2)
Re {A_{\sigma\pi\pi}(s;T)\over 1-2G\Pi_\sigma(s;T)} \cr
&\sum_{c,s}{}'T^{(t,dir)}T^{*(t,exc)}={-g_\pi^4\over 2N_c} \cr}
$$
------------------------------------------------------------------------------
---------------------------------

\vskip 0.2in
\noindent
Figure Captions.

Fig.1.  The diagrams considered in this paper for the hadronization of a
quark--antiquark pair (single lines) into two pions (double lines). The
convention is chosen that they should be read from left to right. Fig.1 a shows
the annihilation of a $q\bar{q}$ via the intermediate formation of a
$\sigma$-meson in the s-channel, while in Fig.1 b, the hadronization proceeds
via quark exchange in the t-channel.

\vskip 0.1in
Fig.2.  The integrated hadronization cross-section $\sigma_q^{had}$
as a function of the c.m.~energy $s$ of the $q\bar{q}$ pair for three
values of the temperature. For comparison, also the integrated elastic
cross-section is shown as a function
of $s$ at a temperature slightly below the critical value.

\vskip 0.1in
Fig.3.  The integrated hadronization cross-section as a function of $T$
for three values of the center of mass energy $s$. Note the (logarithmic)
singularity at the critical temperature $T_c$.

\vfill
\bye